\documentclass[twocolumn,aps,pra,superscriptaddress,showpacs]{revtex4-1}
\usepackage{graphicx,amsmath,amssymb}
\usepackage{dcolumn,fancyhdr}
\usepackage{bm,natbib,mathrsfs}
\usepackage{times}
\usepackage{txfonts}
\usepackage{epstopdf}
\usepackage{latexsym}
\usepackage{epsfig,palatino}
\usepackage{color}
\usepackage{float}

\newcommand{\ket}[1]{\left| #1 \right\rangle}
\newcommand{\bra}[1]{\left\langle #1 \right|}

\newcommand{\nbar}{\overline{n}}

\definecolor{Blue}{rgb}{0,0,1}
\definecolor{Red}{rgb}{1,0,0}
\definecolor{Green}{rgb}{0,1,0}
\definecolor{Purp}{rgb}{.2,0,.2}
\definecolor{white}{rgb}{1,1,1}

\newcommand{\ie}{\textit{i.e.}}

\begin{document}

\title{On the limitations of a measurement-assisted optomechanical route to quantum macroscopicity of superposition states}

\author{Andrew Carlisle} 
\affiliation{Centre for Theoretical Atomic, Molecular and Optical Physics, School of Mathematics and Physics, Queen's University, Belfast BT7 1NN, United Kingdom}
\author{Hyukjoon Kwon}
\affiliation{Center for Macroscopic Quantum Control, Department of Physics and Astronomy, Seoul National University, Seoul, 151-742, Korea}
\author{Hyunseok Jeong}
\affiliation{Center for Macroscopic Quantum Control, Department of Physics and Astronomy, Seoul National University, Seoul, 151-742, Korea}
\author{Alessandro Ferraro} 
\affiliation{Centre for Theoretical Atomic, Molecular and Optical Physics, School of Mathematics and Physics, Queen's University, Belfast BT7 1NN, United Kingdom}
\author{Mauro Paternostro} 
\affiliation{Centre for Theoretical Atomic, Molecular and Optical Physics, School of Mathematics and Physics, Queen's University, Belfast BT7 1NN, United Kingdom}

\date{\today}

\begin{abstract}
Optomechanics is currently believed to provide a promising route towards the achievement of genuine quantum effects at the large, massive-system scale. By using a recently proposed figure of merit that is well suited to address continuous-variable systems, in this paper we analyze the  requirements needed for the state of a mechanical mode (embodied by an end-cavity cantilever or a membrane placed within an optical cavity) to be qualified as macroscopic. We show that, according to the phase space-based criterion that we have chosen for our quantitative analysis, the state achieved through strong single-photon radiation-pressure coupling to a quantized field of light and conditioned by measurements operated on the latter might be interpreted as macroscopically quantum. In general, though, genuine macroscopic quantum superpositions appear to be possible only under quite demanding experimental conditions. 

\end{abstract}

\maketitle

\section{Introduction}

The superposition principle is one of the most distinctive features of quantum mechanics. It is at the basis of phenomena such as entanglement and is at the core for the speed-up of quantum computation and the superiority of quantum communication schemes over their classical counterpart. 

The strenuous experimental efforts produced in the last fifty years have certified the possibility to engineer coherent superpositions in a number of physical systems, from all-optical to solid-state and cold-atom ones. Recent landmarks in such endeavours are embodied by the generation of small-scale Schr\"odigner-cat-like states~\cite{Haroche,JeongNature,Friedman,Kirchmair}. 
Yet, despite the exciting experimental progresses made in this context, we are still facing a rather disappointing lack of tests of the validity of the quantum superposition principle at the genuine meso-/macroscopic scale~\cite{Arndt}.

Notwithstanding such bottlenecks, there is great interest in engineering quantum superpositions of macroscopically distinct states for both technological and foundational reasons. Amongst the various platforms that have been put forward so far to achieve such goal~\cite{Arndt}, a potentially very promising candidate is provided by the framework of cavity optomechanics~\cite{RMP}, which offers unprecedented possibilities to induce strong non-classical features in a massive mechanical system by means of radiation-pressure, combined with perspectives for achieving quantum control at the mesoscopic scale through hybrid settings~\cite{Rogers}. However, a rigorous assessment of such optomechanical route towards macroscopic superpositions is still lacking.

In this paper we investigate the possibilities for the generation of genuinely macroscopic quantum superpositions offered by cavity optomechanics. We analyze both a linear and a quadratic coupling of the intensity of a cavity field with the position of a mechanical oscillator, which are configurations that have been thoroughly explored experimentally~\cite{RMP}. 
We assume both the single-photon optomechanics condition and the intense-driving one. While the first embodies a {\it desideratum} that is currently pursued experimentally, the second adheres very well with current experimental state-of-the-art working regimes.  As a figure of merit for quantum macroscopicity, we employ a recently proposed phase-space measure~\cite{measure}, which is based on a phase-space analysis of the state of harmonic oscillators. So far, various measures for quantum macroscopicity have been proposed~\cite{measure,Leggett, Dur, Shimizu, Bjork, Shimizu2, Cavalcanti, Korsbakken, Marquardt, Korsbakken2, Frowis, Nimmrichter, Sekatski,Yadin,Oudot}. Other than the one chosen for our study~\cite{measure}, the approach suggested by Nimmrichter and Hornberger~\cite{Arndt,Nimmrichter} can also be applied to optomechanical systems. However, their measure is suitable mainly to compare different experimental approaches rather than to compare quantum states per se produced at fixed timings~\cite{Jeong}. The various measures put forward so far appear to capture different aspects and definition of macroscopic quantumness. The one adopted here allows for a fair comparison between states of an optomechanical system achieved by varying the optomechanical coupling strength (at set values of all the other parameters that set a given working point). 
We show that, in the single-photon optomechanical scenario, quantum superpositions that are deemed to be macroscopic in nature by the quantifier that we use can be engineered under conditions of very strong coupling and accurate postselection of the mechanical state following homodyning of the cavity field that drives its motion. 

The remainder of this paper is organised as follows. Sec.~\ref{lin} addresses the case of the coupling between the intensity of a cavity field and the position of a mechanical oscillator. Sec.~\ref{quad} extends such analysis to the coupling to the square of the mechanical system's position. In Sec.~\ref{openlin} we study the effects of both cavity losses and mechanical damping on the results achieved in Sec.~\ref{lin}, thus providing a benchmark for experimentally realistic conditions. Finally, in Sec.~\ref{conc} we draw our conclusions.

\section{Linear coupling configuration}
\label{lin}

Let us start considering the standard description of the radiation-pressure coupling between the field of an optomechanical cavity and a highly reflecting vibrating end-cavity cantilever. The system is described quantum mechanically by the Hamiltonian~\cite{Law}
\begin{equation}
\label{hamiltonian1}
\hat H=\hbar \omega_{0} \hat a^{\dagger} \hat a+ \hbar \omega_{m} \hat b^{\dagger} \hat b-\hbar g \hat a^{\dagger} \hat a \left( \hat b + \hat b^{\dagger} \right)
\end{equation}
where $\omega_{m}$ ($\omega_0$) is the frequency of the mechanical mode (cavity field) and $\hat b, \hat b^\dag$ ($\hat a,\hat a^\dag$) are its bosonic annihilation and creation operators. In Eq.~(\ref{hamiltonian1}), $g=({\omega_{0}}/{L})\sqrt{{\hbar}/({2 m \omega_{m}})}$ is the optomechanical coupling rate. For this first part of our analysis we assume ideal operating conditions --- namely, we neglect losses and describe the dynamics unitarily. This idealisation satisfactorily models a system in the strong single-photon optomechanical coupling regime, \ie~when $g\gg\kappa,\gamma_m$ --- $\kappa$ being the photon loss rate of the cavity field and $\gamma_m$ the mechanical damping rate. While such condition has not yet been met experimentally, significant progress in this sense are currently ongoing~\cite{RMP}. Moreover, such assumption embodies the ``best possible scenario'' and, as such, it allows us to benchmark our analysis. We stress that a study of experimentally realistic conditions is reported later on in this paper.  

\begin{figure}[t]
{\bf (a)}
\includegraphics[width=\columnwidth]{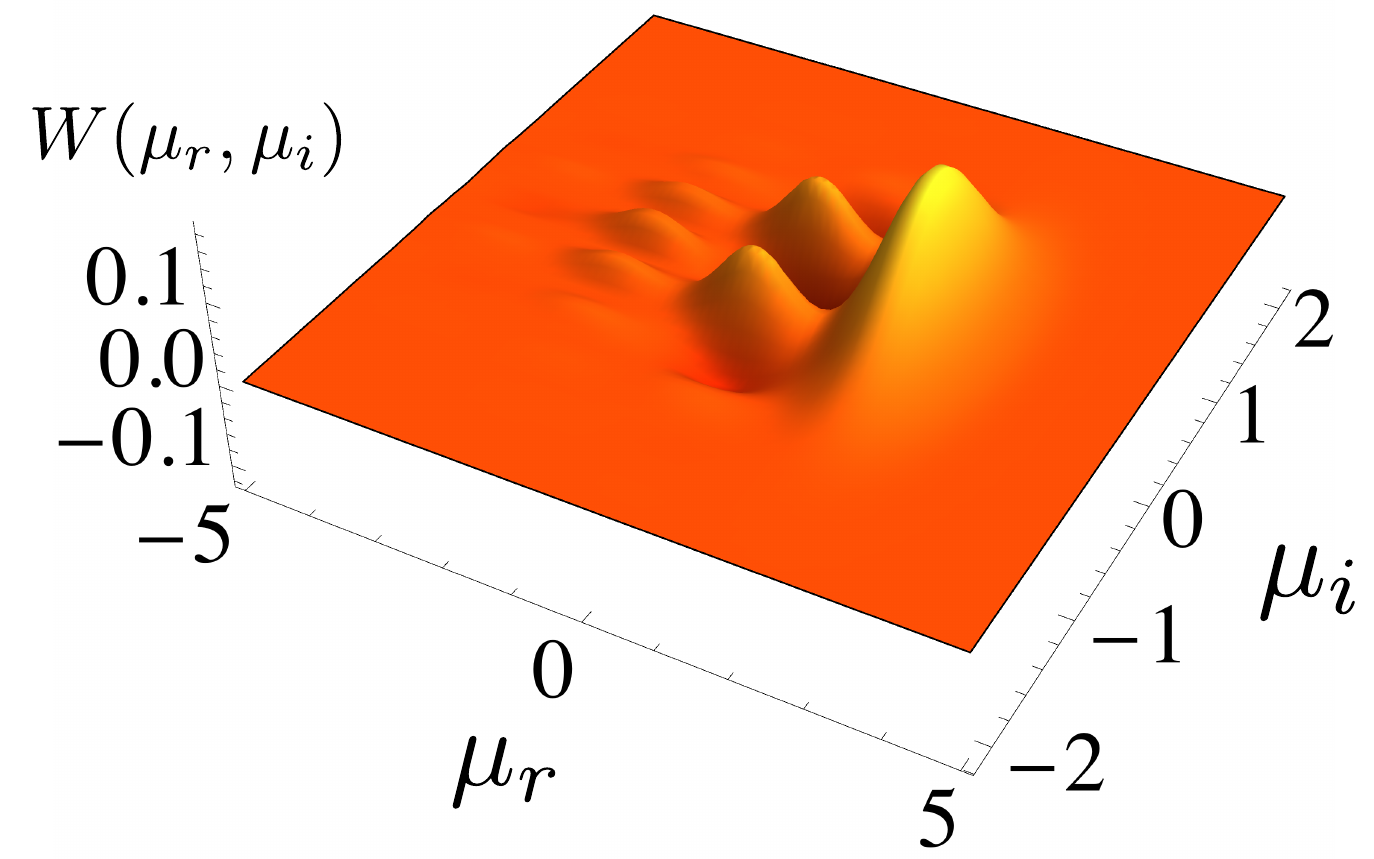}
{\bf (b)}
\includegraphics[width=\columnwidth]{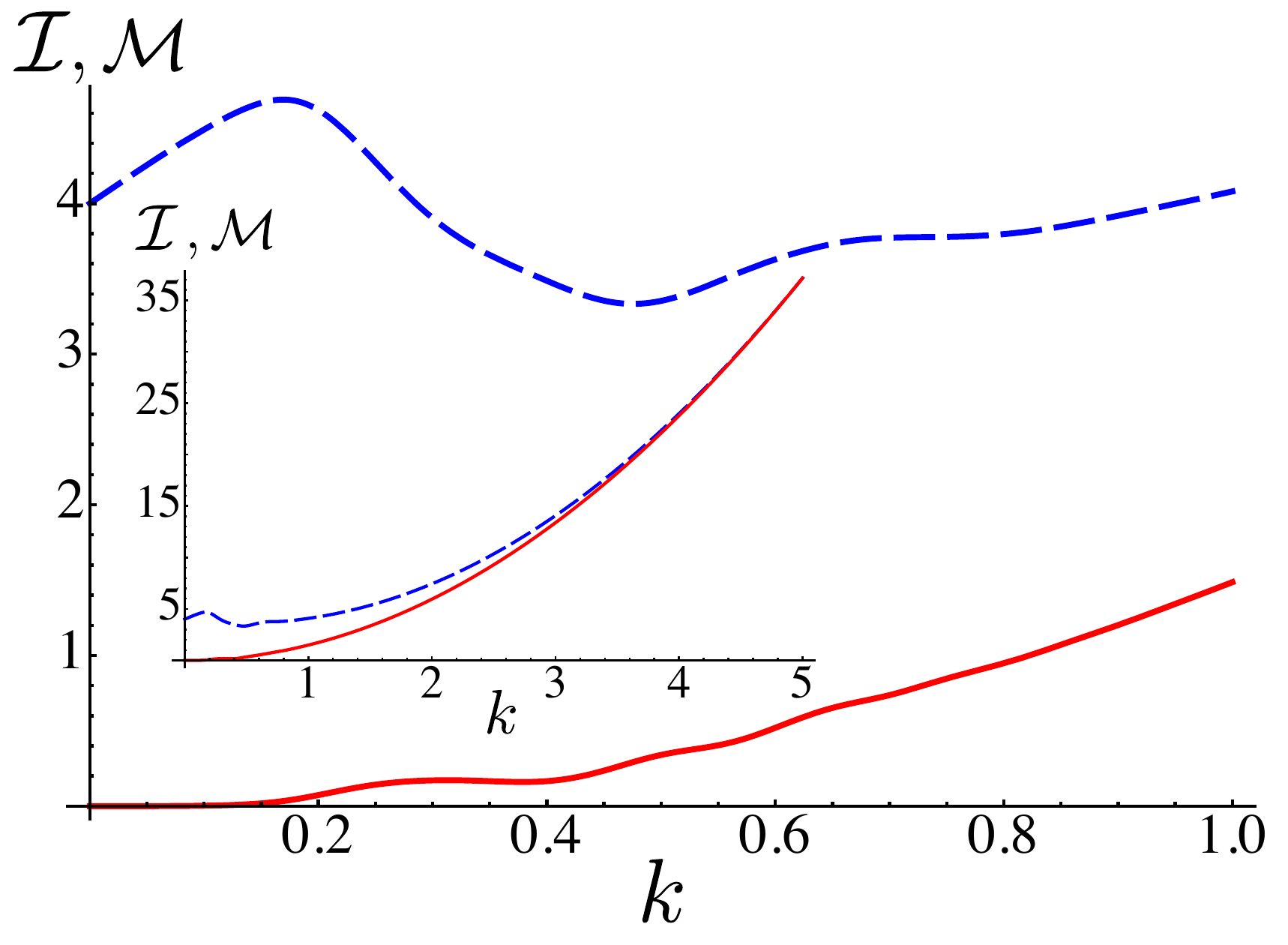}
\caption{(Color online) {\bf (a)} Wigner function of the conditional state of the mechanical mirror achieved by projecting the cavity field into the origin of the phase space. {\bf (b)} Measure of quantum macroscopicity ${\cal I}$ and mean number of phonons ${\cal M}$ in the conditional mechanical state, plotted against the (dimensionless) coupling rate $k$. The inset is for $k\in[0,5]$. In both panels we have taken $T=0$, $\alpha=0.8$, $\beta_0=2$ and $t=\pi$. The Wigner function in panel {\bf (a)} has been evaluated using $k=1$. All the quantities being plotted are dimensionless.}
\label{cat1}
\end{figure}

The time evolution operator generated by $\hat H$ can be put into the form~\cite{system}
\begin{equation}
\label{evolution} 
\hat U(t)=e^{-i r a^{\dagger} a t} e^{i k^{2} (a^{\dagger} a)^{2} \left(t-\mathrm{sin} t\right)}
e^{k a^{\dagger} a ( \eta b^{\dagger} - \eta^{\*} b )} e^{-i b^{\dagger} b t}
\end{equation}
where we have introduced the dimensionless parameters  $t=\omega_m\tau$, $\eta=(1-e^{-it})$, $k={g}/{\omega_{m}}$, and $r={\omega_{0}}/{\omega_{m}}$. Here, $\tau$ is the actual duration of the evolution. The initial state of the system is taken to be $\rho_0=\ket{\alpha}\bra{\alpha}_c\otimes\rho^{\rm th}_{m}$ with $\ket{\alpha}_c$ a coherent state of the cavity field, and $\rho^{\rm th}_m$ a displaced thermal state of the mechanical mode at temperature $T$. The cavity field state is realized by driving the resonator with an external coherent pump, while the mechanical oscillator is assumed to be a displaced state thermalized to the temperature of its phononic environment. Considering the quantum phase-space of the mechanical system and using the $P$ quasi-probability distribution of the latter, we have 
\begin{equation}
\rho^{\rm th}_m=\int P^{\rm th}(\beta)\ket{\beta}\!\bra{\beta}_md^2\beta~~~~~~(\beta\in\mathbb{C})
\end{equation}
with $P^{\rm th}(\beta)=e^{-|\beta-\beta_0|^2/\nbar}/(\pi\nbar)$, $\beta_0$ the amplitude of the mechanical displacement, and $\nbar=(e^{\frac{\hbar\omega_m}{K_B T}}-1)^{-1}$ the mean phonon number ($K_B$ is the Boltzmann constant). The evolved state of the whole system is thus $\int\,P^{\rm th}(\beta)[\hat{U}(t)\ket{\alpha,\beta}\!\bra{\alpha,\beta}_{c,m}\hat{U}^\dag(t)]d^2\beta$ with 
\begin{equation}
\hat U(t)\ket{\alpha,\beta}_{c,m}=e^{-\frac{|\alpha|^2}{2}}\sum_n\frac{\alpha^n}{\sqrt{n!}}e^{ik^2n^2(t-\sin t)}\ket{n,\varphi_n(t)}_{c,m}
\end{equation}
where $\ket{\varphi_n(t)}_m$ are coherent states of the mechanical oscillator with amplitude $\varphi_n(t)=\beta e^{-it}+k n(1-e^{-i t})$. For $T=0$, the analysis in Ref.~\cite{system} revealed that, owing to the large entanglement established between the mechanical mirror and the cavity field at $t=\pi$, a suitable homodyne measurement performed over the field prepares the mirror in a superposition of well-distinguishable coherent states. In particular, by projecting the cavity field into the eigenstate corresponding to the zero of the position quadrature and taking $\alpha=0.8$ and $\beta_0=2$, the conditional state $\ket{\psi(t)}_m$ of the mechanical mirror has an associated Wigner function showing substantial coherences, as revealed by the well-resolved ripples in the plot of Fig.~\ref{cat1} {\bf (a)}. The chosen values of $\alpha$ and $\beta_0$ were demonstrated in Ref.~\cite{system} to be optimal for the production of a cat-like state. The initial state of the system could be prepared by exploiting a competing coupling between the mechanical mode at hand and a second field, addressing it from the back and {\it de facto} cancelling the optomechanical coupling under scrutiny here. This allows for the preparation of a state of the required form~\cite{brunelli}. In fact, the mechanical state resulting from the competition of the two optomechanical coupling mechanisms is very close to the thermal state of its free Hamiltonian~\cite{brunelli}. Turning off the competing coupling for a time $\tau$ would result in the effective switching on of the optomechanical interaction for the needed time interval. 

To address how macroscopic such quantum superposition state is, we now employ the measure introduced in Ref.~\cite{measure}, which is fundamentally based on the extent of such ripples and thus evaluate the quantity 
\begin{equation}
\label{measure}
{\cal I}=\mathrm{Max}\left[0,\frac{1}{2 \pi} \int \mathrm{d}^2 \xi (\vert \xi \vert^2-1) \vert \chi_m(\xi) \vert^2\right]
\end{equation}
where $\chi_m(\xi)={}_m\!\bra{\psi(t)}\hat D_m(\xi)\ket{\psi(t)}_m$ is the Weyl characteristic function of the conditional mechanical state state, $\hat D_m(\zeta)$ is the displacement operator, and $\xi\in\mathbb{C}$. It can be shown that ${\cal I}\le{\cal M}$, where ${\cal M}={\rm Tr}[\hat b^\dag\hat b~\rho_m(t)]$ is the mean phonon number at time $t$ and $\rho_m(t)$ is the corresponding reduced state of the mechanical mode. As discussed in Ref.~\cite{measure}, a non-zero value of ${\cal I}$ is an indicator of quantum macroscopicity. Needless to say, the larger the value of ${\cal I}$, the more pronounced is the macroscopic character of the state under scrutiny. A fully consistent framework for the study of macroscopicity is still lacking: at present, various proposals for the quantification of the macroscopic character of the state of a quantum system have been put forward~\cite{measure,Leggett, Dur, Shimizu, Bjork, Shimizu2, Cavalcanti, Korsbakken, Marquardt, Korsbakken2, Frowis, Nimmrichter, Sekatski,Yadin,Oudot}. Yet, a systematic and rigorous approach to the definition of macroscopic quantumness has only recently been presented~\cite{Yadin2015}, highlighting the need for objective criteria that any {\it bona fide} measure should satisfy. The quantity ${\cal I}$ proposed in Ref.~\cite{measure} and used throughout this paper combines handiness of calculation [cf. Eq.~\eqref{measure}], as it relies on the easily accessible Wigner function of a given state, and a clear physical interpretation, being linked directly to the rate of change of the purity of a given quantum system~\cite{measure}.

\begin{figure*}[t!]
\centering
\hskip0.5cm{\bf (a)}\hskip5cm{\bf (b)}\hskip5cm{\bf (c)}
\includegraphics[width=0.67\columnwidth]{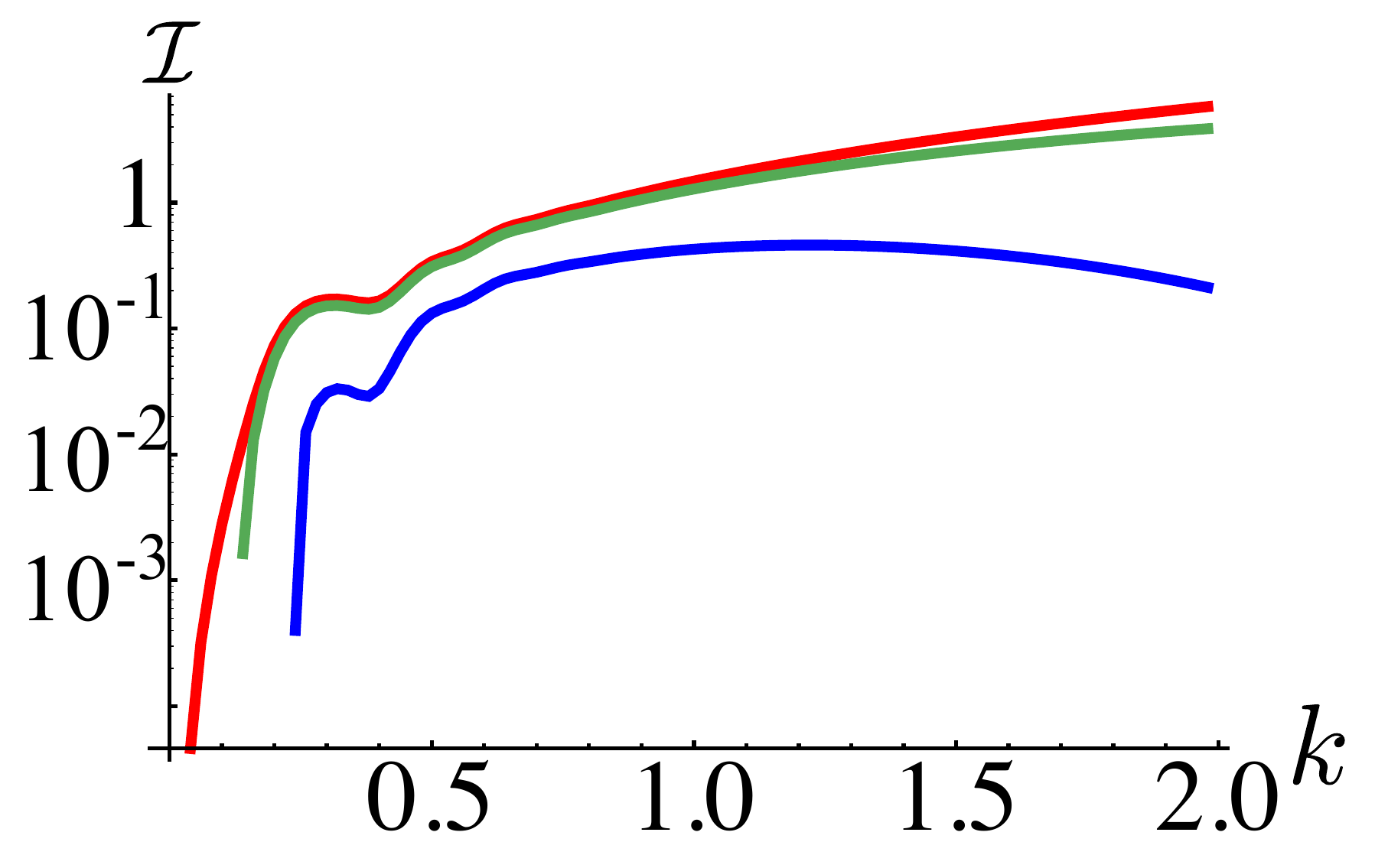}\includegraphics[width=0.67\columnwidth]{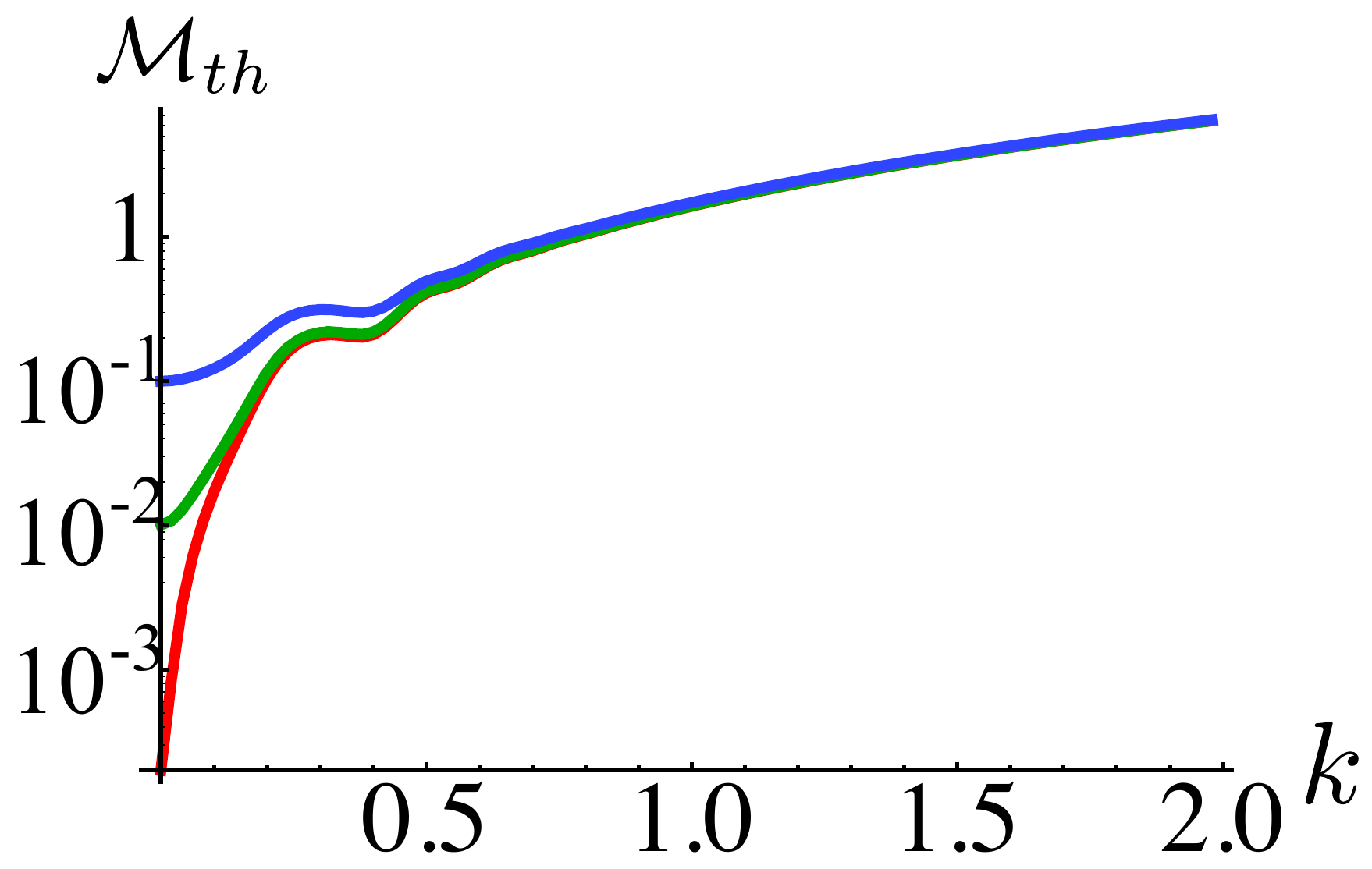}\includegraphics[width=0.67\columnwidth]{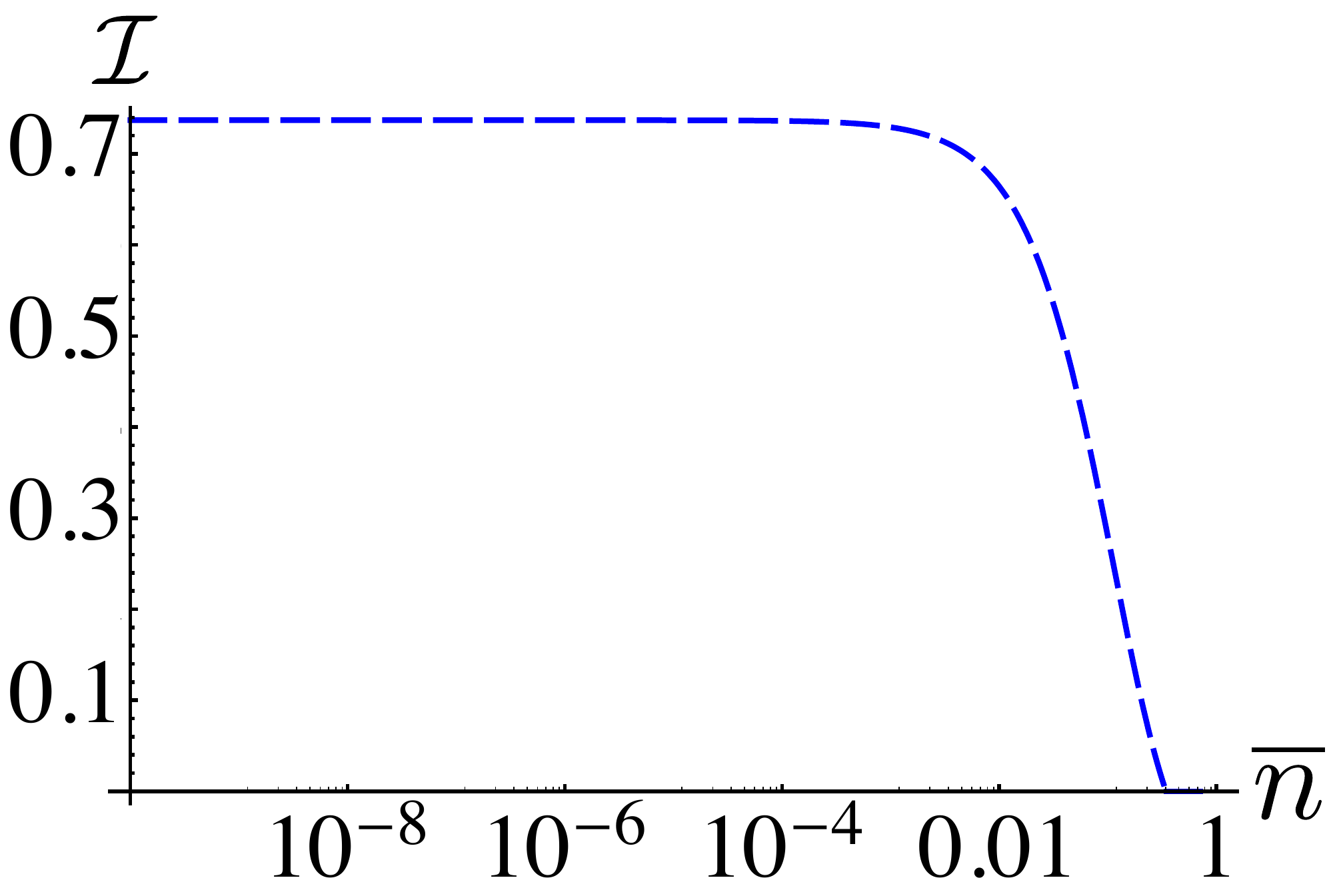}
\caption{(Color online) {\bf (a)} [{\bf (b)}] Measure of quantum macroscopicity [Average number of phonons] for the conditional state of a mechanical mode obtained taking $\alpha=0.8$, $\beta_0=0$ and projecting the cavity field onto the origin of its phase space.  From bottom to top [top to bottom] curve, we have $\nbar=0.1,10^{-2}$ and $10^{-4}$, respectively. {\bf (c)} Maximum degree of macroscopicity in the conditional state of an end-cavity mechanical oscillator, plotted against the mean occupation number $\nbar$ of the initial mechanical state. We have taken $k=0.7$ and $\alpha=0.8$. All the quantities being plotted are dimensionless.} 
\label{FP1}
\end{figure*}

In Fig.~\ref{cat1} {\bf (b)} we show the behavior of the quantum macroscopicity measure ${\cal I}$ and the mean number of phonons ${\cal M}$ against the dimensionless coupling rate $k$ for $T=0$ (so that the mechanical system is initially prepared in a coherent state of amplitude $\beta_0$), and the same working point used to plot the Wigner function of Fig.~\ref{cat1} {\bf (a)}. As a quantitative benchmark, we compare the value of ${\cal I}$ achieved in our case with what would be obtained using a Schr\"odinger cat state of the form $\ket{{\cal C}}={\cal N}(\ket{\alpha_0}+\ket{-\alpha_0})$ with $\ket{\pm\alpha_0}$ a coherent state of amplitude $\alpha_0\in{\mathbb R}$ and ${\cal N}$ a normalisation factor. This class of states saturate the upper bound set to ${\cal I}$, i.e. they are such that ${\cal I}={\cal M}$ and  embody a significant milestone. Quantitatively, a $\ket{\cal C}$ state achieves a value of quantum macroscopicity ${\cal I}_{\cal C}$ for an amplitude $\alpha_0$ satisfying the condition $\tanh(\alpha^2)={\cal I}_{\cal C}/\alpha_0^2$. At $\beta_0=2$ with $\alpha=0.8$, $T=0$, and $k=1$, which would imply a coupling strength of the order of the mechanical frequency, the conditional mechanical state achieved by projecting the state of the cavity field onto the origin of the phase space achieved a degree of quantum macroscopicity ${\cal I}\simeq1.49$. A state $\ket{{\cal C}}$  with such value of quantum macroscopicity would require coherent states of amplitude $\alpha_0\simeq1.27$, which would hardly qualify the corresponding cat-like state as macroscopic. Only values of the coupling strength $k\ge10$ would correspond to degree of quantum macroscopicity comparable with sizeable cat-like states, as amplitudes $\alpha_0\ge10$ would be correspondingly required. Moreover, as the single-photon optomechanical coupling strength depends on $1/\sqrt{m}$, mechanical systems of large mass give rise to small values of $k$. This is intuitive, as more photons would be needed to put in motion an oscillator of large mass. By fixing the density of the material used to fabricate the mechanical oscillator, then, similar considerations hold for its size.  

In the foregoing analysis, the zero temperature assumption for the mechanical mode is rather stringent. Although experimental success in the cooling of a mechanical mode all the way down to its ground-state have been reported recently~\cite{cooling}, this is still a very challenging task, in particular if combined with the need for control over the cavity field as well. We have thus relaxed such assumption to study the impact that a non-zero temperature would have on the degree of macroscopicity of the system. As our results turn out to be independent of the initial displacement of the mechanical mirror, in our calculations we have set $\beta_0=0$ and computed the thermal-averaged quantities $\chi_{th}(\zeta)=\int \mathrm{d}^{2}\beta P^{\rm{th}}(\beta) \chi_m(\xi)\vert_{\beta_0=0}$, which in turn allows us to evaluate the corresponding macroscopicity measure ${\cal I}$ and the thermal occupation number ${\cal M}_{th}=\int \mathrm{d}^{2} \beta P^{\rm{th}}(\beta) {\cal M}$. The results are reported in Fig.~\ref{FP1} {\bf (a)} and {\bf (b)}, where it can be seen that, by smearing out coherences, the increasingly thermal character of the mechanical state reduces the degree of quantum macroscopicity, rendering it no longer a monotonic function of the coupling strength $k$ (this is due to the fact that ${\cal M}_{th}$ increases with $k$, while ${\cal I}$ does not). We have thus performed a scaling analysis of the macroscopicity measure at a value of the coupling strength that is large enough to induce strong single-photon optomechanical effects, yet somehow foreseable, and against the mean thermal occupation number $\nbar$ which is related to temperature [cf. Fig.~\ref{FP1} {\bf (c)}], finding that the highest temperature allowing for a significant value of ${\cal I}$ for a mechanical oscillator with $\omega_m=1$ MHz is as small as $1~\mu$K. Above such threshold, although coherences might be established, the latter are not strong enough to qualify the quantum mechanical state as macroscopic accordingly to the measure considered here.

\section{Quadratic coupling configuration} 
\label{quad}

We now modify the configuration of our thought experiment and consider the setting where a partially reflecting structure is placed within the volume of a fixed-boundary cavity. Under proper arrangements, such configuration gives rise to an overall quartic Hamiltonian of the form~\cite{biancofiore} 
\begin{equation}
\hat H=\hbar \omega_{0} \hat a^{\dagger} \hat a+ \hbar \omega_{m} \hat b^{\dagger} \hat b-\hbar g \hat a^{\dagger} \hat a ( \hat b + \hat b^{\dagger} )^{2}\;,
\end{equation}
a possibility that has been demonstrated experimentally in Ref.~\cite{Thompson}. As a result of such coupling, the mechanical system undergoes squeezing by a degree that depends on the number of photons in the cavity field. For a cavity initially prepared in a coherent state $\ket{\alpha}$ and the membrane in its vacuum state (i.e. for a ground-state cooled mechanical oscillator), it is possible to prove that the time-evolved state reads~\cite{quad}
\begin{equation}
\vert \psi(t) \rangle=\sum_{n=0}^{\infty} c(n,t)  \vert n \rangle_{c}\vert \zeta(n) \rangle_{m}
\end{equation}
where $c(n,t)={\alpha^{n} e^{-\frac{1}{2} (\vert \alpha \vert^{2}+i \eta)}}/{\sqrt{n !}}$, $\ket{\zeta(n)}$ is a single-mode squeezed state of squeezing degree $\zeta(n)= i e^{i\eta}{{\rm arcsinh}}\left( \frac{2kn \, \mathrm{sin} \chi \, t}{\chi} \right)$,
and $\eta=\text{arctan} \left( \frac{ \rho }{2 \chi}\tan \chi t \right)$ with $\chi=\sqrt{1+4 k n}$, $\rho=-2(1+2kn)$ and, as before, $k=g/\omega_m$.

\begin{figure*}[t!]
\centering
{\bf (a)}\hskip4cm{\bf (b)}\hskip4cm{\bf (c)}\hskip4cm{\bf (d)}
\includegraphics[width=0.5\columnwidth]{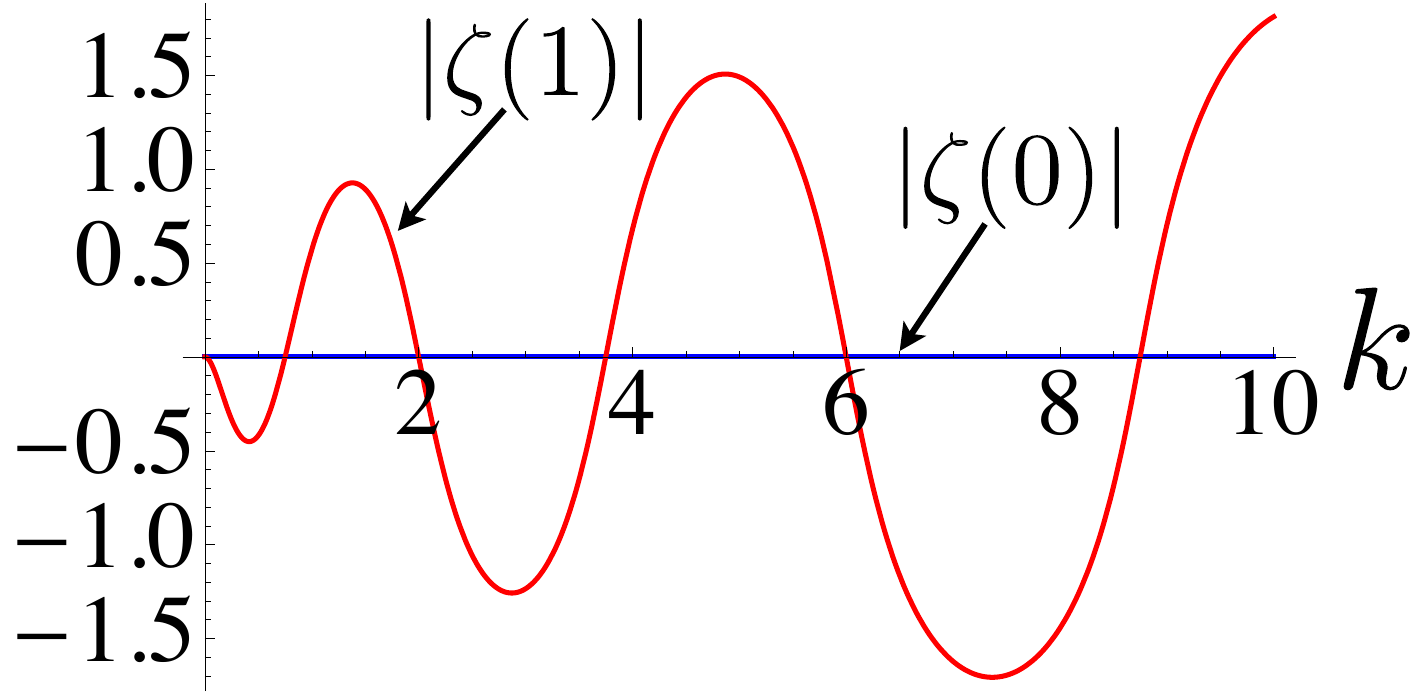}~~\includegraphics[width=0.5\columnwidth]{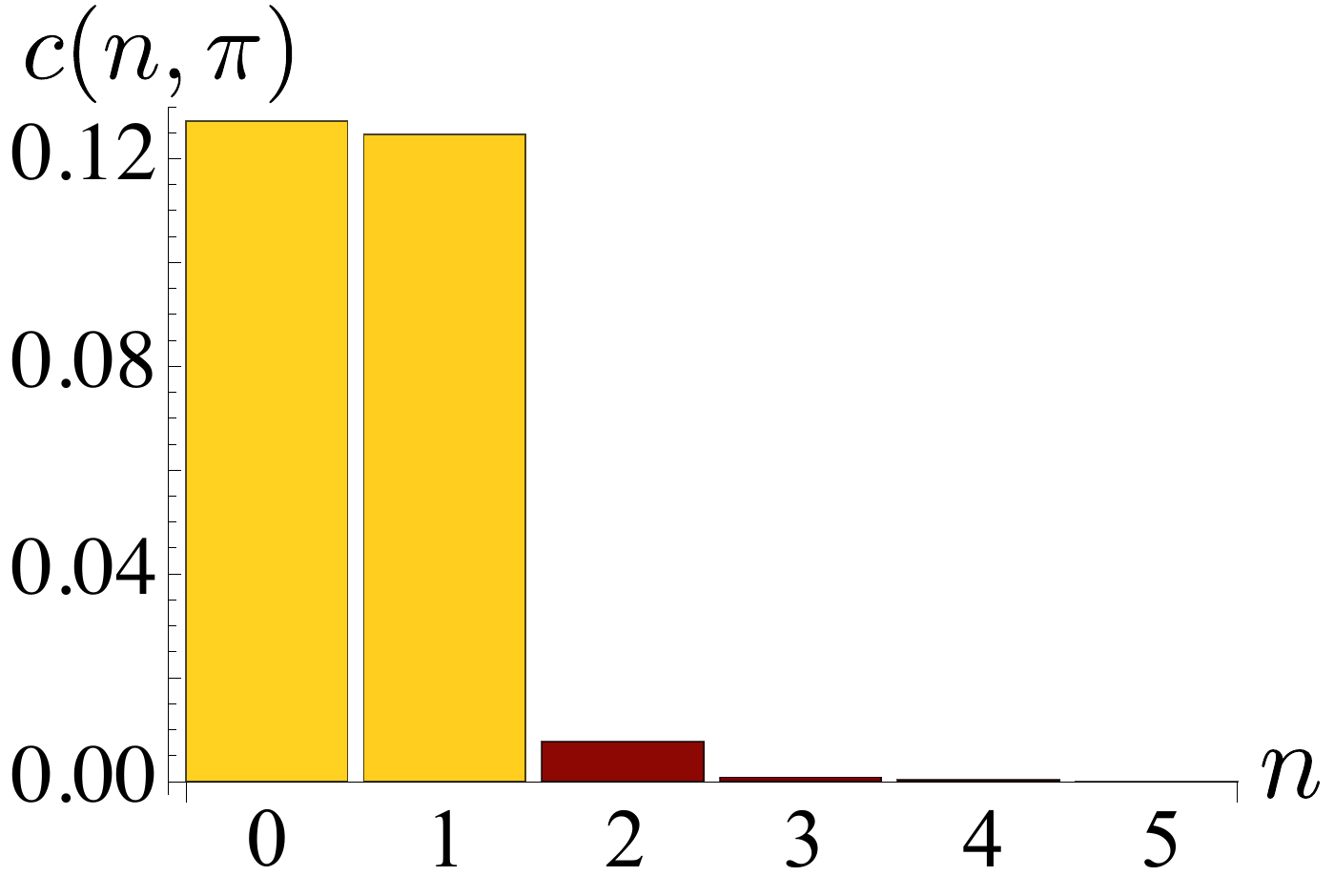}~~\includegraphics[width=0.5\columnwidth]{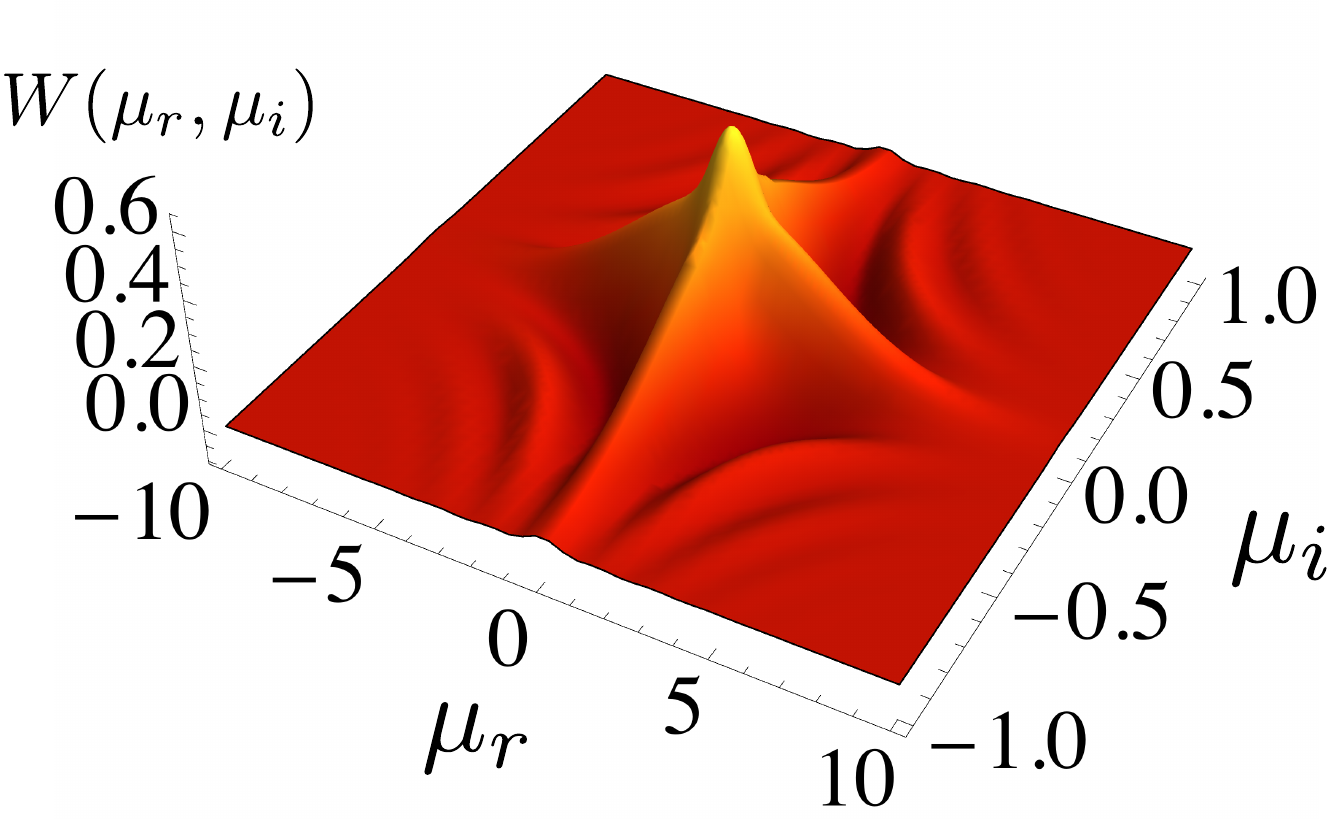}\includegraphics[width=0.5\columnwidth]{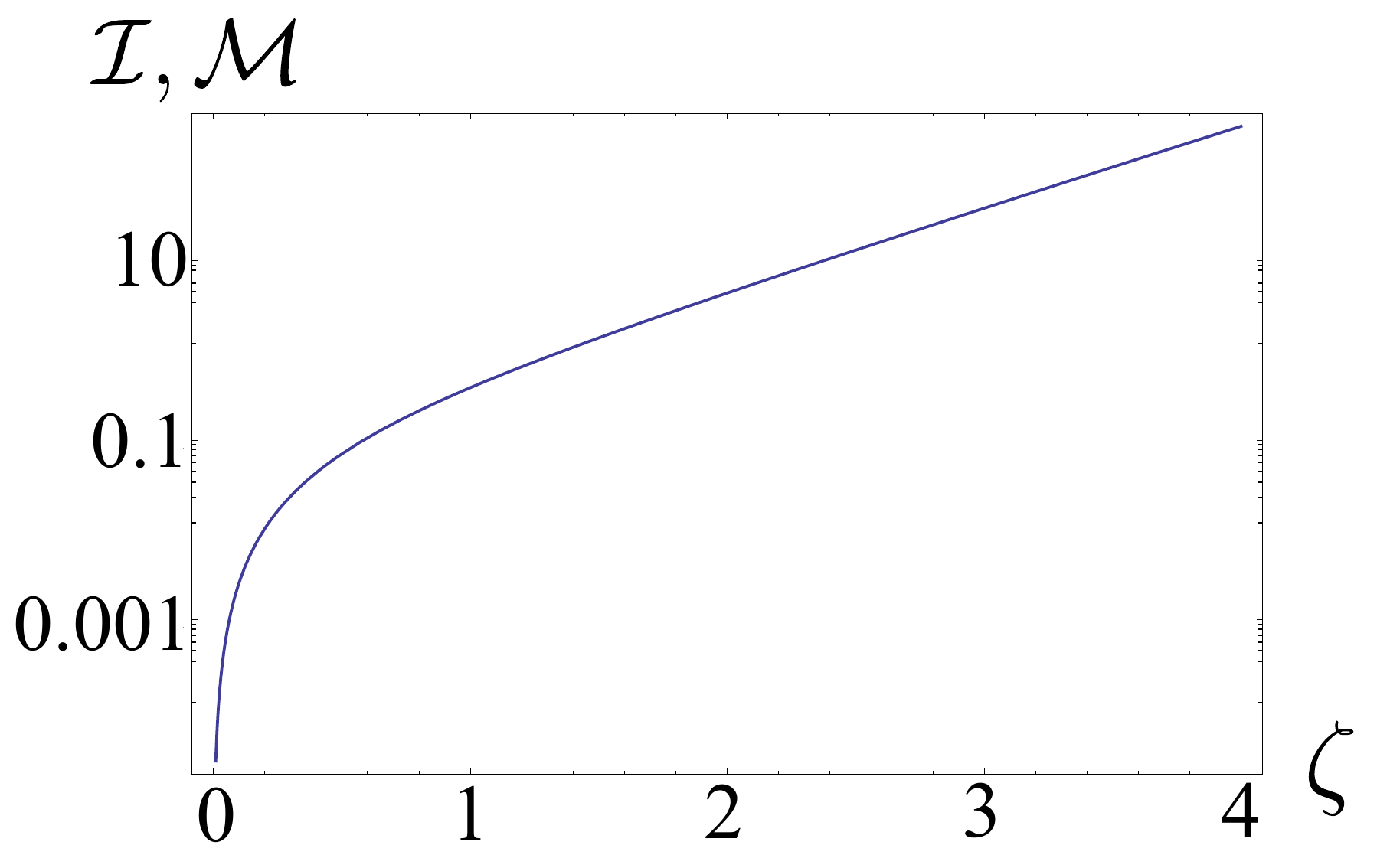}
\caption{(Color online) {\bf (a)} Behavior of the degree of squeezing $\zeta(n)$ for $n=0$ and $n=1$ against the dimensionless coupling strength $k=g/\omega_m$ at $\tau=\pi$. {\bf (b)} Distribution of the probability amplitudes in the state of the membrane achieved when taking $\alpha=0.7$, $x=1$, $k=1$, $t=\pi$. {\bf (c)} Wigner function associated with the state $\ket{\psi'_\pi}\simeq{\cal N}(\ket{0}+\ket{\zeta(1)})$ of the mechanical system for a value of $k$ such that $|\zeta(1)|=2$. {\bf (d)} Measure of macroscopicity and mean phonon number plotted against the degree of squeezing $\zeta=|\zeta(1)|$. Only at large degrees of squeezing is the state of the mechanical system macroscopically coherent. All the quantities being plotted are dimensionless.} 
\label{squeezebehavior}
\end{figure*}

As for the previous configuration, we seek to enforce mechanical coherence by performing a suitable projective measurement on the light. Following the same lines highlighted before, we homodyne the field, postselecting the outcome corresponding to a projection onto the the quadrature eigenstate $\ket{x}$. The correspondingly reduced state of the mechanical system reads (for $\alpha\in\mathbb{R})$
\begin{equation}
\label{statequad}
\vert \psi'_{t} \rangle 
  =N \sum_{n=0}^{\infty} \frac{\alpha^n}{\sqrt[4]{\pi}} \frac{e^{-\frac{x^{2}}{2}-\frac{ \alpha^{2}}{2}+i\eta}}{\sqrt{2^n} n !} H_{n}(x) \vert \zeta(n) \rangle,
\end{equation}
where $N$ is a normalization factor and $H_n(x)$ are Hermite polynomials of order $n$. We now aim at identifying the conditions that make $\ket{\psi'_t}$ a coherent superposition of two highly distinguishable squeezed states. As the energy of a squeezed state is directly proportional to the degree of squeezing, we would like the latter to be sufficiently large. For this, we need to understand the behavior of the degree of squeezing $|\zeta(n)|$, which is plotted against the interaction strength $k$ and for $n=0,1$ in Fig.~\ref{squeezebehavior} {\bf (a)}. A quick numerical optimisation over the other parameters entering Eq.~(\ref{statequad}) shows that for $x=1$, $\alpha=0.7$ and $t=\pi$, we have $c(0,\pi)\simeq c(1,\pi)$, with $c(n\ge2,\pi)$ being negligibly small [cf. Fig.~\ref{squeezebehavior} {\bf (b)}], and thus $\ket{\psi'_\pi}\simeq{\cal N}(\ket{0}+\ket{\zeta(1)})$ with ${\cal N}^2=\sqrt{\cosh\zeta(1)}/[2+2\sqrt{\cosh\zeta(1)}]$. The Wigner function associated with such state is shown in Fig.~\ref{squeezebehavior} {\bf (c)}. An explicit calculation yelds that 
\begin{equation}
{\cal I}={ \cal M}= \frac{\sqrt{\cosh\zeta(1)}(\sinh\zeta(1))^2}{2(1+\sqrt{\cosh \zeta(1)})}.
\end{equation}
implying that the measure of macroscopicity saturates its upper bound. However, as said, this is not sufficient to claim for the macroscopic character of the quantum superpostion and the actual values of ${\cal I}$ and ${ \cal M}$ need to be considered.
Fig.~\ref{squeezebehavior} {\bf (d)} displays both of them against the degree of squeezing $\zeta=|\zeta(1)|$. Clearly, they become significant only for values of $\zeta\gtrsim2$, which in turn requires $k\gtrsim17$. Such conditions are currently technologically prohibitive, thus forcing us to conclude that, even for the quartic hamiltonian coupling under scrutiny here, no macroscopic superposition of mechanical states of motion can be engineered without either dramatically improving the technology or adopting a different strategy to enforce quantum superpositions.

\section{Linear-coupling open-system case} 
\label{openlin}

Although it is instrumental in providing insight into the difficulties of producing genuinely macroscopic mechanical superposition states, the approach above retains elements of ideality that set it apart from the current experimental reality. Therefore, in order to adhere more closely to the situations currently addressed in an optomechanics laboratory, here we analyze the degree of macroscopicity of the mechanical state produced upon implementing a strategy analogous to the one described above when all sources of noise are included and the assumption of large single-photon optomechanical rate is relaxed. 

A quantitative analysis performed by making use of a quantum Langevin equations approach~\cite{linear} shows that, upon postselection of the mechanical state following the projection of the cavity field onto the origin of the phase space, no macroscopic character is displayed. Let us sketch the formal steps towards such conclusions.

\begin{figure}[t!]
\centering
\includegraphics[width=\columnwidth]{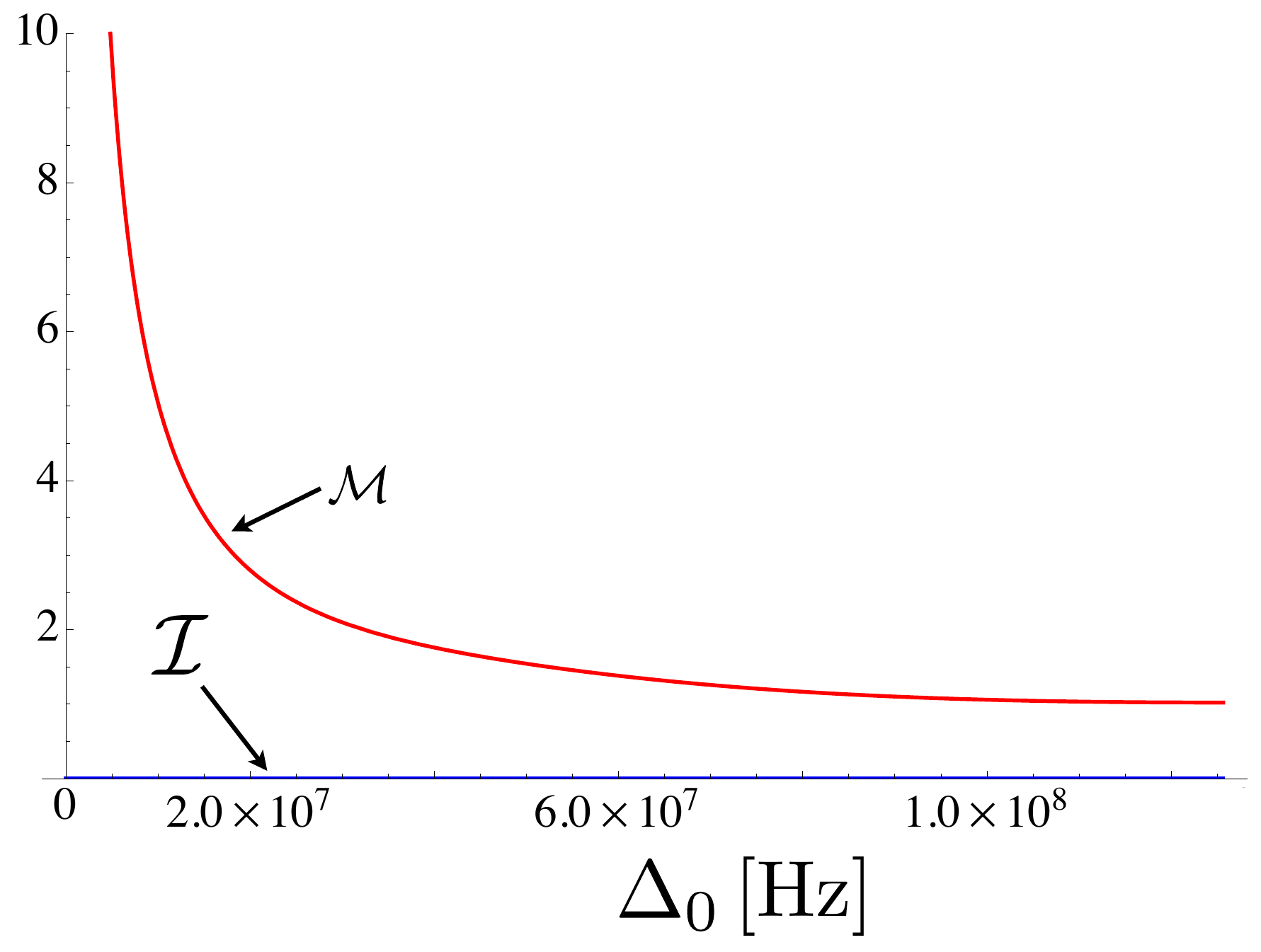}
\caption{(Color online) We show the measure of macroscopicity ${\cal I}$ and the mean phonon number ${\cal M}$ against the cavity-pump detuning $\Delta_0$ for $\gamma_m/2\pi=100$ Hz, $\omega_m/2\pi=10$ MHz, $\kappa=88$ MHz, $\epsilon=6\times10^{12}$ Hz and a mechanical system of $5$ng mass in a cavity of $1$ mm length. The external pump has frequency $\omega_L/2\pi=3.7\times10^{14}$ Hz and the operating temperature is taken to be the rather optimistic value of $0.4$ K. Both ${\cal I}$ and ${\cal M}$ are dimensionless.} 
\label{badnews}
\end{figure}

The dynamical equation for a driven optomechanical cavity in the case of a linear coupling of the cavity field to the position of the mechanical oscillator read
\begin{equation}
\label{langevin}
\begin{aligned}
 \dot{\hat q}&=\omega_{m} \hat p,\\
 \dot{\hat p}&=-\omega_{m} \hat q - \gamma_{m} \hat p - g \hat a^{\dagger} \hat a + \hat \epsilon,\\
\dot{\hat a}&=-(\kappa+i \Delta_{0}) \hat a - i g \hat a \hat q + E+\sqrt{2\kappa}\hat a^{\mathrm{in}}.
\end{aligned}
\end{equation}
Here $\hat{q}=(\hat b^\dag+\hat b)/\sqrt{2}$ and $\hat{p}=i(\hat b^\dag-\hat b)/\sqrt{2}$ are the dimensionless position and momentum quadratures for the mirror, $a^{\mathrm{in}}$ is the vacuum radiation input noise to the cavity, $E$ is the strength of the coupling between the pump (at frequency $\omega_L$ and the cavity), $\hat \epsilon$ is a noise operator that accounts for the Brownian motion undergone by the mechanical oscillator (which is in contact with a bath of phononic modes at temperature $T$), and $\Delta_{0}$ is the pump-cavity detuning. For an intense pumping field, the cavity field and mechanical system fluctuate quantum mechanically around classical stationary values of their amplitude and position, respectively. This allows us to consider only the quantum fluctuations $\delta{u}$ of the operators $\hat u=(\hat q,\hat p,\hat a,\hat a^\dag)$ entering Eqs.~\eqref{langevin}. This leaves us with a new set of equations whose solution, achieved as illustrated in Refs.~\cite{pater,linear,covar}, allows for the analytic calculation of the steady-state covariance matrix of the system~\cite{covar,pater1}, which is defined as
\begin{equation}
V_{ij}=\frac{1}{2}(\langle \delta u_{i}(\infty) \delta u_{j}(\infty)+\delta u_{j}(\infty) \delta u_{i}(\infty) \rangle).
\end{equation}
In this context, one can easily incorporate the effects that conditional measurements over the state of the light field have on the mechanical state. For instance, the covariance matrix of the mechanical system 
\begin{equation}
V_m=
\begin{pmatrix}
V_{11}&V_{12}\\
V_{21}&V_{22}
\end{pmatrix}
\end{equation}
is modified by a homodyne measurement performed over the cavity field as~\cite{scheel}
\begin{equation}
V'_m=V_m-V_c(\Pi V_f\Pi)^{-1}V^T_c
\end{equation}
with 
\begin{equation}
\Pi=
\begin{pmatrix}1&0\\0&0\end{pmatrix},\quad V_{f}=
\begin{pmatrix}
V_{33}&V_{34}\\
V_{43}&V_{44}
\end{pmatrix},\quad V_{c}=
\begin{pmatrix}
V_{13}&V_{14}\\
V_{23}&V_{24}
\end{pmatrix}.
\end{equation}
In turn, this allows for the evaluation of the Weyl characteristic function associated with the conditional state of the mechanical system as $\chi(\zeta)=e^{-\frac{1}{2} \zeta {V'_m} \zeta^{T}}$ with $\zeta$ the complex phase space variable, and thus the calculation of the measure of macroscopicity. A typical plot of both ${\cal I}$ and ${\cal M}$ against the experimentally adjustable parameter $\Delta_0$ is presented in Fig.~\ref{badnews} for a set of parameters very close to what can be currently achieved in quantum optomechanics laboratories.

Clearly, the state resulting from homodyning the field that has interacted with the mechanical system shows no macroscopic quantumness, when the figure of merit embodied by ${\cal I}$ is considered. While the mean number of phonons populating the state of the mechanical system is non-zero for all values of $\Delta_0$ considered, the measure of macroscopicity is identically null. We have performed  similar study for the case of quadratic coupling to the position of the mechanical oscillator, finding again no macroscopically quantum character. 

\section{Conclusions} 
\label{conc}

We have explored the possibility to enforce macroscopic quantum coherence in the state of a mechanical system driven by light. By analyzing two optomechanical coupling models currently at the core of extensive experimental investigations, and armed with the tools provided by the measure proposed in Ref.~\cite{measure}, we have demonstrated the necessity of large or ultra-large single-photon optomechanical coupling strengths and postselection for the sake of producing significant degrees of macroscopicity. 

We believe that, despite the somehow negative character of such surprising results, useful information can nevertheless be gathered from our analysis. In particular, our findings would point the current experimental efforts towards the direction aimed at achieving truthful macroscopic mechanical states. Such endeavours will have to be focused on the achievement of large single-photon optomechanical couplings, and the implementation of conditional strategies able to "extract" coherence of a macroscopic nature from the state of the mechanical systems used in typical optomechanical experiments. In this respect, our future investigations will address the possibilities offered by pulsed optomechanics~\cite{pulse}, which implements effective homodyne measurements on the mechanical device and could be useful in generating sizeable macroscopic quantum states. 

\acknowledgments

We thank H. Ulbricht, V. Vedral, and A. Xuereb for valuable discussions on the topic of this manuscript. This work was supported by the UK EPSRC (through grant EP/G004579/1), the John Templeton Foundation (grant ID 43467), the Northern Ireland Department of Employment and Learning. H.K. and H.J. acknowledge the National Research Foundation of Korea (NRF) grant funded by the Korea government (MSIP) (No. 2010-0018295).

\end{document}